\documentclass[a4paper,fleqn]{cas-sc}

\usepackage[authoryear,longnamesfirst]{natbib}
\usepackage[ruled, lined, linesnumbered, commentsnumbered, longend]{algorithm2e}
\usepackage{amsmath}
\usepackage{hyperref}
\def\tsc#1{\csdef{#1}{\textsc{\lowercase{#1}}\xspace}}
\tsc{WGM}
\tsc{QE}
\tsc{EP}
\tsc{PMS}
\tsc{BEC}
\tsc{DE}

\begin{document}
\let\WriteBookmarks\relax
\def\floatpagepagefraction{1}
\def\textpagefraction{.001}

\shorttitle{K-CROSS}

\shortauthors{Guoyang Xie et~al.}

\title [mode = title]{K-Space-Aware Cross-Modality Score for Quality Assessment of Synthesized Neuroimages}                      



%

\author[1,2]{Guoyang Xie}[]

\cormark[1]

\ead{guoyang.xie@ieee.org}


\credit{Conceptualization, Methodology, Data Curation, Validation, Visualization, Writing - original draft \& editing}

\affiliation[1]{organization={Southern University of Science and Technology},
    addressline={Department of Computer Science}, 
    city={Shenzhen},
    postcode={518055}, 
    country={China}}

\affiliation[2]{organization={CATL},
    addressline={Department of Intelligent Manufacturing}, 
    city={Ningde},
    postcode={352000}, 
    country={China}}

\author[3]{Jinbao Wang}[]
\cormark[1]
\ead{wangjb@szu.edu.cn}
\affiliation[3]{organization={Shenzhen University},
    addressline={National Engineering Laboratory for Big Data System Computing Technology}, 
    city={Shenzhen},
    postcode={518060}, 
    country={China}}
\credit{Conceptualization, Methodology, Data Curation, Investigation, Writing - original draft \& editing}
\author[4]{Yawen Huang}[]

\ead{bear\_huang@126.com}
\cormark[1]

\credit{Conceptualization, Methodology, Investigation, Writing - review \& editing, Supervision}

\affiliation[4]{organization={Tencent Youtu Lab},
    addressline={Jarvis Research Center}, 
    city={Shenzhen},
    postcode={510310}, 
    country={China}}

\author%
[5]
{Jiayi Lyu}
\ead{lyujiayi21@mails.ucas.ac.cn}
\credit{Data Curation, Visualization, Writing - review \& editing}

\affiliation[5]{organization={University of Chinese Academy of Sciences},
    addressline={the School of Engineering Science}, 
    city={Beijing},
    postcode={100049}, 
    country={China}}

\author%
[1]
{Feng Zheng}
\cormark[2]
\ead{f.zheng@ieee.org}
\credit{Supervision, Project administration, Funding acquisition, Writing - review \& editing}

\author%
[4]
{Yefeng Zheng}
\ead{yefengzheng@tencent.com}
\credit{Supervision, Writing - review \& editing}
\author%
[6]
{Yaochu Jin}
\ead{jinyaochu@westlake.edu.cn}
\credit{Supervision, Writing - review \& editing}

\affiliation[6]{organization={Westlake University},
    addressline={the School of Engineering}, 
    city={Hangzhou},
    postcode={310030}, 
    country={China}}

\cortext[cor1]{Contributed Equally}
\cortext[cor2]{Corresponding Author}



\begin{abstract}
The problem of how to assess cross-modality medical image synthesis has been largely unexplored. The most used measures like PSNR and SSIM focus on analyzing the structural features but neglect the crucial lesion location and fundamental k-space speciality of medical images. To overcome this problem, we propose a new metric K-CROSS to spur progress on this challenging problem. Specifically, K-CROSS uses a pre-trained multi-modality segmentation network to predict the lesion location, together with a tumor encoder for representing features, such as texture details and brightness intensities. To further reflect the frequency-specific information from the magnetic resonance imaging principles, both k-space features and vision features are obtained and employed in our comprehensive encoders with a frequency reconstruction penalty. The structure-shared encoders are designed and constrained with a similarity loss to capture the intrinsic common structural information for both modalities. As a consequence, the features learned from lesion regions, k-space, and anatomical structures are all captured, which serve as our quality evaluators. We evaluate the performance by constructing a large-scale cross-modality neuroimaging perceptual similarity (NIRPS) dataset with 6,000 radiologist judgments. Extensive experiments demonstrate that the proposed method outperforms other metrics, especially in comparison with the radiologists on NIRPS. To forster reproducibility and accessibility, the source codeis uploaded to the website: \href{https://github.com/M-3LAB/K-CROSS}{https://github.com/M-3LAB/K-CROSS} 
\end{abstract}


\begin{highlights}
\item  We propose a new metric, called K-CROSS, to evaluate the quality of the synthetic data based on all the structural information, k-space feature shift, and lesion area. This multidimensional quantification indication enables K-CROSS to achieve more precise results than other metrics that only consider natural images.
\item To properly verify the effectiveness of our K-CROSS, we construct a large-scale and multi-modal neuroimaging perceptual similarity (NIRPS) dataset, which includes 6,000 assessments from radiologists.
\item K-CROSS achieves highly competitive results based on the judgments from radiologists on NIRPS, which can be treated as a general evaluation metric for various purposes of medical image synthesis.
\end{highlights}


\maketitle

\section{Introduction}

PSNR~\citep{HuynhThu2008ScopeOV}, SSIM~\citep{Wang2004ImageQA}, and MAE~\citep{Chai2014RootMS} are the most commonly used evaluation metrics in cross-modality magnetic resonance imaging (MRI) synthesis works. However, these metrics are inappropriate to a certain degree, considering that they are based on natural images and naturally ignore the inherent properties of MRI data. In general, the quality of neuroimage can be assessed by the content (\textit{i.e.}, lesion region), frequency space, and structure details. Although MAE, PSNR and SSIM are effective in assessing image quality, they are ineffective as a neuroimage metric, because they only focus on the structural details in the pixel space. Therefore, it is important to find a new way to measure how good the cross-modality neuroimage synthesis is. To forster reproducibility and accessibility, the source code is uploaded to .

\begin{figure}[t]
	\centerline{\includegraphics[width=1\linewidth]{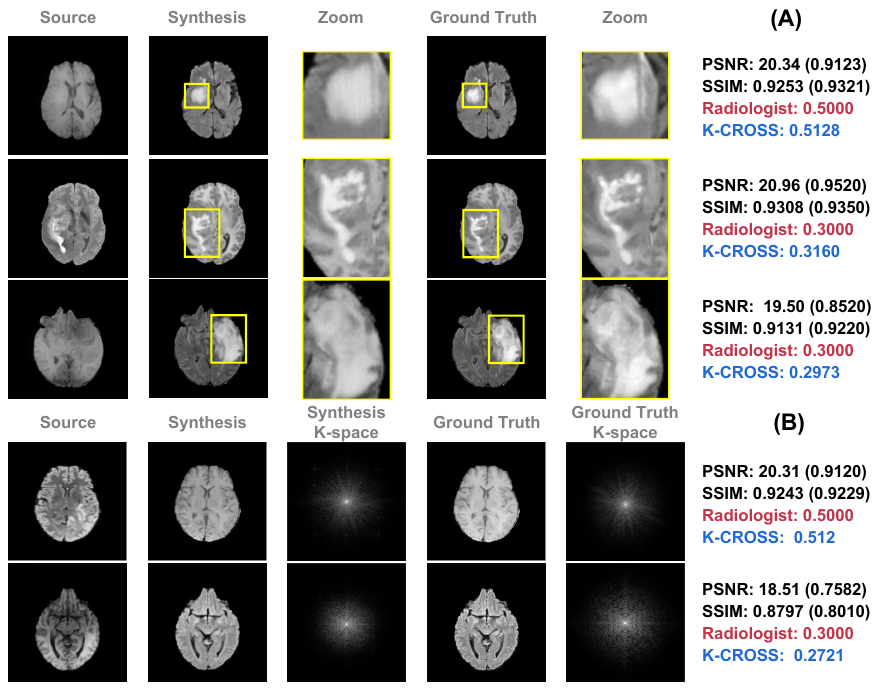}}
	\caption{K-CROSS vs. PSNR vs. SSIM. The first and second columns on the left represent the source and synthesized target modality data, respectively. The zoom indicates the modality-specific tumor region, which is provided by the pre-trained multi-modality neuroimage segmentation network. The numbers on the right represent PSNR, SSIM, radiologist score, and our K-CROSS value. PSNR, SSIM, and K-CROSS are rescaled (to 1.0000) for comparison with the radiologist's score. In terms of lesion region and k-space measurement, K-CROSS is more compatible with the radiologist's score than PSNR and SSIM.}
    \label{fig:motivation}
\end{figure}

Empirically, the content details of neuroimages, particularly the texture and brightness, are disregarded by either PSNR or SSIM. Instead, radiologists pay more attention to the lesion regions, since the usefulness of analyzing pathology and human cognitive functions. The purpose of K-CROSS is to fully reflect the lesion region by introducing a cross-modality neuroimage segmentation network which has already been trained to precisely forecast the tumor location. The prediction mask (\textit{i.e.,} tumor region) is fed into the proposed tumor encoder to extract features. The proposed tumor loss function improves the extracted feature to capture more essential texture details and brightness information. In Fig.~\ref{fig:motivation} (A), we can observe that the content of the synthesis neuroimage does not align with the target modality neuroimage. Though PSNR and SSIM scores are the highest for the synthesized ones, they only evaluate the structure details without taking the content into account. By contrast, K-CROSS is reliable in exploring neuroimaging perceptual similarity (NIRPS) for the synthesized results.

Besides, PSNR and SSIM are unable to account for differences in the k-space between the synthesized images and the target modality data, whereas K-CROSS can. The fundamental difference between MRI and natural images, as seen from the standpoint of imaging principles, is the basis of MR image reconstruction. The Fourier transformation, often known as the "k-space" in MRI, is a mathematical concept that calculates various frequencies mixed into the received signal of all spins. It forms the basis for all image reconstruction in MRI. Therefore, we believe that the proposed metric can estimate the distance between MRIs in both k-space and pixel space. In k-space, the sophisticated K-CROSS encoder can capture the invariant modal-specific feature, where the frequency loss can be used to further enhance the complicated encoder.  When a k-space shift occurs, as seen at the bottom of Fig.~\ref{fig:motivation} (B), K-CROSS is more stable in accordance with the radiologist’s score, which is able to measure the gap in k-space between the synthesized neuroimage and the corresponding ground truth.

To constrain the structural features that are extracted by the shared structure encoder from both the source modality and the target modality, we set up a cross-modality similarity loss function, as the entire structure information between the source and the target modality neuroimaging data is very similar. PSNR and SSIM, on the other hand, only assess the input image, which limits their capacity to recognize the structural details that the source modality and the target modality share.

Our contributions can be summarized as follows:
\begin{itemize}
    \item We propose a new metric, called K-CROSS, to evaluate the quality of the synthetic data based on all the structural information, k-space feature shift, and lesion area. This multidimensional quantification indication enables K-CROSS to achieve more precise results than other metrics that only consider natural images.
   
    \item To properly verify the effectiveness of our K-CROSS, we construct a large-scale and multi-modal neuroimaging perceptual similarity (NIRPS) dataset, which includes 6,000 assessments from radiologists.
    
    \item K-CROSS achieves highly competitive results based on the judgments from radiologists on NIRPS, which can be treated as a general evaluation metric for various purposes of medical image synthesis.
    
\end{itemize}

The rest of this paper is organized as follows: Section~\ref{sec:related-work} presents a literature review on image quality assessment and GAN-based assessment methods. Section~\ref{sec:proposed-method} explains the proposed algorithm K-CROSS in detail. In addition, a large-scale multi-modal neuroimaging perceptual similarity (NIRPS) dataset is constructed in Section~\ref{sec:nirps-dataset}. Section~\ref{sec:experiment} presents comprehensive experimental evaluations while Section V draws the conclusion and limitation of the current work.

\section{Related Work} \label{sec:related-work}

\subsection{Image Quality Assessment}
Image quality assessment (IQA) can be divided into two categories. One is fully referenced IQA, and the other is non-referenced IQA~\citep{Mittal2013MakingA, Moorthy2011BlindIQ, Bosse2016ADN, Talebi2018NIMANI}. IQA with all references refers to estimating the quality of natural images with references: SSIM~\citep{Wang2004ImageQA}, MS-SSIM~\citep{Wang2003MultiscaleSS} and FSIM~\citep{Zhang2011FSIMAF}, focus more on image structure specifics. Specifically, FSIM builds up a novel feature similarity index according to the phase congruence and image gradient magnitude, while PSNR focuses on edge estimation for the synthesized images. Most of them~\citep{Lin2018HallucinatedIQANI, Ren2018RAN4IQARA, Lim2018VRIN, Zhang2021UncertaintyAwareBI} use low-level features for evaluation. LPIPS~\citep{Zhang2018TheUE} is the first work that uses a high-level feature for fully referenced IQA in light of the popularity of deep learning. Estimating the synthesized image quality without a reference (ground truth) is known as non-referenced IQA~\citep{Mittal2013MakingA, Moorthy2011BlindIQ, Bosse2016ADN, Talebi2018NIMANI}. RankIQA~\citep{Liu2017RankIQALF} is the mainstream for non-referenced IQA. Considering the limited size of IQA, Liu~\textit{et al.} propose a Siamese network to rank images and their distorted ones. The Siamese network's knowledge (ranking result) can be transferred to a conventional neural network, whose function is to assess the quality of a single image.  Since K-CROSS requires a reference image for evaluation, it belongs to a fully referenced IQA. However, few public data in the medical imaging community could be used to train for the learning-based fully referenced IQA methods. The NIRPS dataset, the first extensive neuroimaging perceptual similarity dataset with radiologists' labels is constructed. As for fully referenced IQA methods, K-CROSS is, therefore, able to use the supervised training methods.

\subsection{GAN Assessment}
The existing sample-based methods~\citep{Xu2018AnES} have been proposed to access GAN performance, like Kernel MMD~\citep{Gretton2012OptimalKC}, Inception Score~\citep{Salimans2016ImprovedTF}, Mode Score~\citep{Che2017ModeRG} and FID~\citep{Heusel2017GANsTB}. The classical approach is to compare the log-likelhood of generative models. But this approach cannnot accurately indicate the quality of synthesized image. In other words, a model can achieve high likelihood, but low image quality, and conversely, low image quality, and conversely. As for Inception score~\citep{Salimans2016ImprovedTF}, it computes the KL divergence between the conditional class distribution and the marginal class distribution over the generated data. However, IS does not capture intra-class diversity, which is insensitive to the prior distribution over labels. Among them, the most popular metric is FID. Heusel~\textit{et al.}~\citep{Heusel2017GANsTB} use InceptionV3~\citep{Szegedy2016RethinkingTI} to extract the features from the real and synthetic neuroimaging data, and then compute the differences in the features between them. However, the majority of them are created in the pixel space and ignore the lesion region and k-space, which are the fundamental elements of MR image properties. In this regard, K-CROSS considers the underlying MR imaging principle as well as the difference between the neuroimages of the source and target modality.

\subsection{Cross-Modality Medical Image Synthesis}
Existing medical image-to-image translation~\citep{jiang2019synthesize,ren2021segmentation,kong2021breaking} has demonstrated their considerable research and clinical analysis potential. Of these methods, supervised GANs are still the mainstream for cross-modality neuroimaging data synthesis~\citep{Wang2018LocalityAM,Dar2019ImageSI,Sharma2020MissingMP,Yu2020SampleAdaptiveGL,Yu20183DCB,Zuo2021DMCFusionDM}. For instance, Elad~\textit{et al.}~\citep{elad2012sparse} provide a concise overview of sparse and redundant representation modeling and identify ten important future research directions for sparse coding. Rubinstein~\textit{et al.}~\citep{rubinstein2010dictionaries} explicate the dictionary acquisition process through mathematical models. Gao~\textit{et al.}~\citep{gao2012laplacian} develop a hypergraph Laplacian matrix to retain the local information of the training samples, thereby enhancing the discriminative capacity of the learned dictionary. However, synthesizing in a supervised manner requires paired data for training, which is difficult to implement in practice. To solve this problem, both semi-supervised and unsupervised methods are then launched to eliminate the need of paired data. \citep{Guo2021AnatomicAM} leverage a lesion segmentation network as a teacher to guide the generator by using unpaired training data. \citep{Shen2021MultiDomainIC} and \citep{Zhou2021AnatomyConstrainedCL} also utilize the high-level tasks to guide the cross-modality image synthesis. Huang~\textit{et al.}~\citep{Huang2020SuperResolutionAI, Huang2020MCMTGANMC} make full use of unpaired cross-modality data and project them into a common space. The attributed features from the common space bring great helpful to synthesize the missing target modality data. \citep{Li2022AGF} employ the dual-domain attention mechanism to extract highly discriminative features on the lesion area. Yu~\textit{et al.}~\citep{Yu2021MouseGANGM} present a similar work with the method shown in~\citep{Huang2020MCMTGANMC}. However, the authors pay more attention to the mouse brain dataset. Chen~\textit{et al.}~\citep{Chen2020UnsupervisedBC} propose a more concise idea, i.e., the source modality and the target modality share their feature encoder. Jiao~\textit{et al.}~\citep{Jiao2020SelfSupervisedUT} also extract and map features into the common space using different modalities. Moreover, the authors in~\citep{Jiao2020SelfSupervisedUT} design a new cross-modal attention module for fusion and propagation. Zeng~\textit{et al.}~\citep{Zeng2019HybridGA} use two models; one of which is the 3D generator network, and the other is the 2D discriminator. The authors utilize the result from the 2D discriminator treated as a weak label to supervise the 3D generator, such that the generator's output can be closer to the output of CT. Yang~\textit{et al.}~\citep{Yang2021AUH} design a unified generator for MRI synthesis. \citep{Wu2023AGGNAG} propose a MRI oriented novel attention-based glioma grading network into mutli-scale feature extraction process, which aims to promote the synergistic interaction among different modality information. \citep{kong2021breaking} introduce a new I2IT model called RegGAN, which converts the unsupervised I2IT task into a supervised I2IT with noisy labels. K-CROSS serves as a general metrics for various levels of supervision of cross-modality medical image synthesis.

\begin{figure*}[t]
         \centerline{\includegraphics[width=1\linewidth]{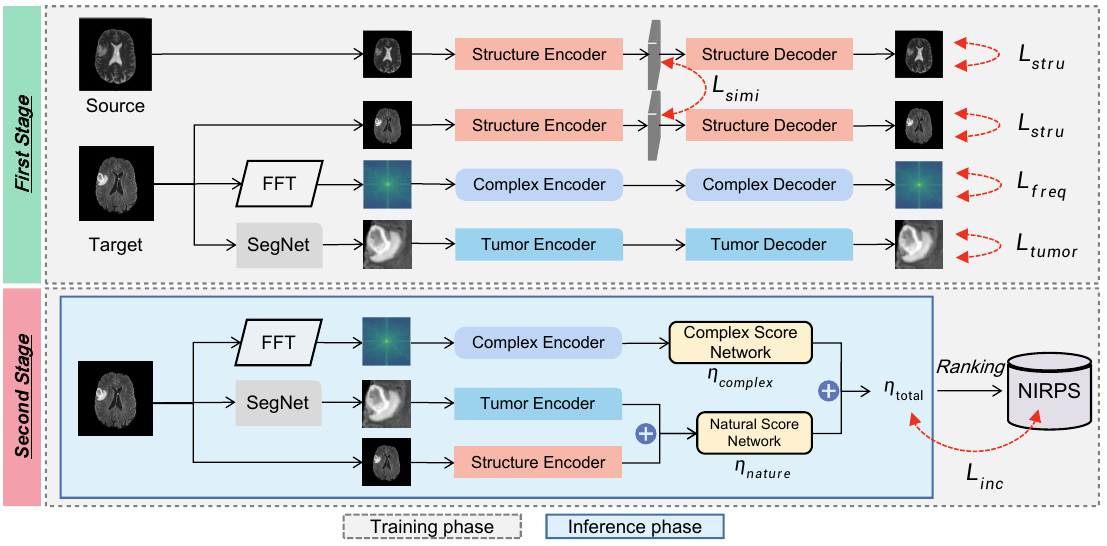}}
	\caption{Flowchart of our proposed K-CROSS. 
	\textbf{First stage input:} For reference neuroimage (source modality) and its query neuroimage (the target modality), the private portion is indicated by a blue box. The neuroimage's structural feature is captured by the structure encoder. The reconstruction specifics are evaluated using $L_{stru}$. With a similarity loss $L_{simi}$, the presentation of the shared structure is maintained. The k-space feature to a modality is captured by Complex U-Net and optimized by $L_{freq}$. SegNet gets the mask of the tumor region from neuroimages that are specific to a given modality. The parameters of the off-the-shelf SegNet are not updated during the training phase, such as nnUnet~\citep{Isensee2020nnUNetAS}, TransUNet~\citep{Chen2021TransUNetTM}, and SwinUNet~\citep{Cao2021SwinUnetUP}. The tumor encoder learns how to represent the tumor mask region, particularly texture details and the level of brightness. The quality of the SegNet-reconstructed tumor region is constrained using the loss $L_{tumor}$. 
	\textbf{First stage output:} The private tumor encoder, the private complex encoder and the shared structure encoder for both modalities. \textbf{Second stage input:} The input are the query modality neuroimage and the modality-specific tumor encoder, complex encoder and the shared structure encoder from the first stage. The two main components of K-CROSS are $\eta_{complex}$ and $\eta_{nature}$. The complex score network yields $\eta_{complex}$, whereas the natural score network yields $\eta_{nature}$. The output of the tumor encoder and the structure encoder are combined as the input of the natural score network. The average score for $\eta_{complex}$ and $\eta_{nature}$ is $\eta_{total}$. For $\eta_{total}$, K-CROSS uses a straightforward regression model during the training phase, with labels taken from the NIPRS dataset.
    \textbf{Inference:} The input are the query modality neuroimage and the modality-specific tumor encoder, complex encoder and the shared structure encoder from the first stage. The output score is $\eta_{total}$.
	}
 \label{fig:K-CROSS}
\end{figure*}

\section{Proposed Method} \label{sec:proposed-method}

\subsection{Preliminary}

\subsubsection{K-Space Representations}
The spatial frequencies of an MR picture are represented in k-space by a matrix of numbers. Despite MR images and k-space having the identical dimension, in practice, each point $(k_{x}, k_{y})$ in k-space represents the spatial frequency and phase information about each pixel in the MR image rather than corresponding to a specific pixel value. By contrast, every pixel in the MR image maps to a point in k-space. As a result, we transform MRI into k-space using the 2D discrete Fourier transform:
\begin{equation} \label{eq:freq_formulation}
    F (u, v) = \sum_{x=0}^{M-1}\sum_{y=0}^{N-1} f(x, y) e^{-i2\pi (\frac{ux}{M} + \frac{vy}{N})},
\end{equation}

where the MR image size is  $M \times N$, $(x,y)$ is the MRI's pixel coordinate, $(u,v)$ is its spatial coordinate in k-space, $F(u,v)$ is its complex frequency value, and $e$ and $i$ stand for the Euler's number and the imaginary unit, respectively. We concentrate on the real and imaginary components of $F(u,v)$. According to~(\ref{eq:freq_formulation}), we rewrite $F(u, v)$ as follows:
\begin{equation}
    F(u, v) = R(u, v) + I(u, v)i = a + bi,
\end{equation}
where the imaginary and real parts of $F(u, v)$ are $I(u, v)$ and $R(u, v)= a$, respectively. Furthermore, we introduce two key k-space concepts. Here, the amplitude can be defined as:
\begin{equation}
    \left| F(u, v) \right| = \sqrt{R(u,v)^{2} + I(u,v)^{2}} = \sqrt{a^{2} + b^{2}}.
\end{equation}
The amplitude is a measure of how strongly a 2D wave reacts to an MR image. We typically visualize k-space using the amplitude.
\begin{equation}\label{eq:phase}
     \angle F(u,v) = arctan\left ( \frac{I(u,v)}{R(u,v)} \right ) = arctan \left( \frac{b}{a} \right).
\end{equation}
The peak shit distance between two 2D sinusoidal waves of the same frequency is referred to as a phase. The phase is the second concept, which is defined in~\eqref{eq:phase}.

\begin{figure*}[t]
	\centerline{\includegraphics[width=\linewidth]{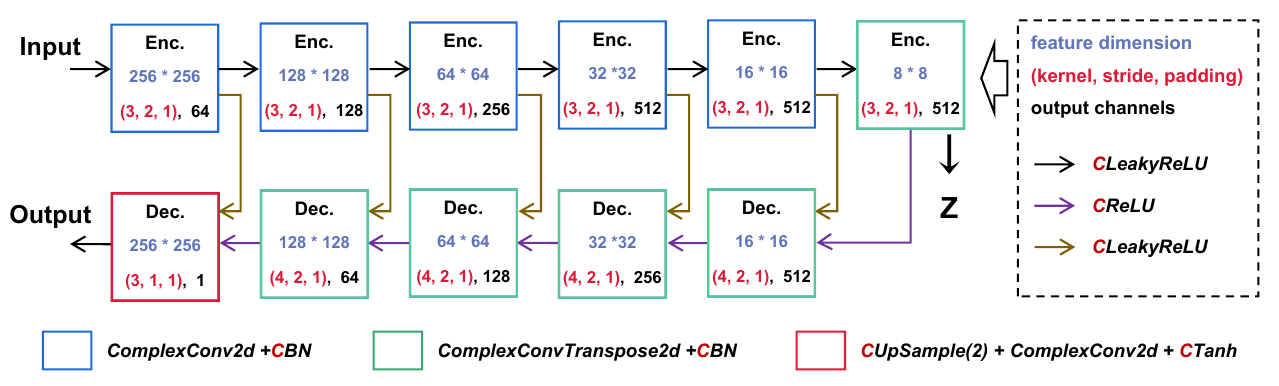}}
	\caption{The architecture of complex encoder and decoder. The encoder consists of Complex Conv2d in Section~\ref{sec:complex_conv} and Complex BatchNorm in Section~\ref{sec:complex_bn}. The decoder contains ComplexConvTranspose2d and Complex BatchNorm. The ComplexConvTranspose2d is similar to ComplxeConv2d except for the convolution operator. The middle part includes Complex Upsample in Section~\ref{sec:complex_up}, Complex Tanh in Section~\ref{sec:complex_tanh} and Complex Conv2d.
	}\label{fig:complex_branch}
\end{figure*}

\begin{figure}[htbp]
	\centerline{\includegraphics[width=0.75\linewidth]{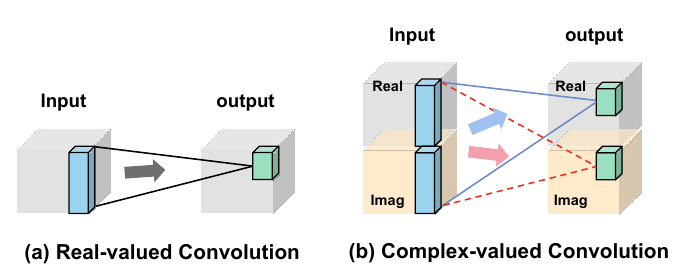}}
	\caption{Illustration of complex convolution.
	}\label{fig:complex_convolve}
\end{figure}

\subsubsection{Complex Convolution}\label{sec:complex_conv}
The complex-valued convolution~\citep{Trabelsi2018DeepCN} is different from the real-valued convolution. Given a complex-valued convolution filter $\textbf{\textit{W}} = \textbf{\textit{A}} + i\textbf{\textit{B}}$ with real-valued matrices $\textbf{\textit{A}}$ and $\textbf{\textit{B}}$. The operation is expressed as follows:
\begin{equation}
    \textbf{W} \ast \textbf{h} = (\textbf{A} \ast \textbf{x} - \textbf{B} \ast \textbf{y}) + i(\textbf{B} \ast \textbf{x} + \textbf{A} \ast \textbf{y}).
\end{equation}
The visualization can be found in Fig.~\ref{fig:complex_convolve}.

\subsubsection{Complex Leaky RELU}
It applies separate Leaky RELUs~\citep{Maas2013RectifierNI} on both the real part $\textbf{\textit{R(z)}}$ and the imaginary part $\textbf{\textit{Im(z)}}$ of a complex-valued, which is defined as:
\begin{equation}
    \mathbb{C}LeakyRELU = LRELU(\textbf{\textit{R(z)}}) + i \ast LRELU(\textbf{\textit{Im(z)}}),
\end{equation}

\subsubsection{Complex RELU}
\begin{figure}[ht]
	\centerline{\includegraphics[width=0.75\linewidth]{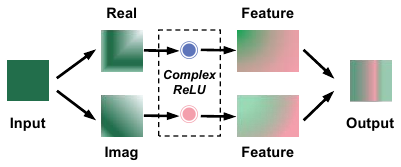}}
	\caption{Illustration of complex operator. 
	}\label{fig:complex_op_style}
\end{figure}

It applies separate RELUs~\citep{Agarap2018DeepLU} on both the real part $\textbf{\textit{R(z)}}$ and the imaginary part $\textbf{\textit{Im(z)}}$ of a complex-valued, which is defined as:
\begin{equation}
    \mathbb{C}RELU = RELU(\textbf{\textit{R(z)}}) + i \ast RELU(\textbf{\textit{Im(z)}}). 
\end{equation}
The visualization is given in Fig.~\ref{fig:complex_op_style}.

\subsubsection{Complex Tanh}\label{sec:complex_tanh}
Complex Tanh applies separate tanh activation~\citep{kalman1992tanh} on both the real part $\textbf{\textit{R(z)}}$ and imaginary part $\textbf{\textit{Im(z)}}$ of a complex-valued, which is defined as:
\begin{equation}
    \mathbb{C}Tanh = Tanh(\textbf{\textit{R(z)}}) + i \ast Tanh(\textbf{\textit{Im(z)}}). 
\end{equation}

\subsubsection{Complex BatchNorm}\label{sec:complex_bn}
As described in Cogswell~\textit{et al.}~\citep{Cogswell2016ReducingOI}, complex-valued batch normalization could be separately applied into the imaginary part and real part, which could reduce the risk of over-fitting. The detail operation is defined as:
\begin{equation}
    \mathbb{C}BN = BN(\textbf{\textit{R(z)}}) + i \ast BN(\textbf{\textit{Im(z)}}).
\end{equation}
\subsubsection{Complex Upsample}\label{sec:complex_up}
complex-valued upsample algorithm is able to be separately applied to the real part and imaginary part, which is defined as:
\begin{equation}
    \mathbb{C}Upsample = Upsample(\textbf{\textit{R(z)}}) + i \ast Upsample(\textbf{\textit{Im(z)}}).
\end{equation}

\subsection{Architecture}
The two stages of training for K-CROSS are depicted in Fig~\ref{fig:K-CROSS}. Encoder-decoder architecture is primarily used in the first stage of K-CROSS to train the tumor encoder, complex encoder, and structure encoder. The complex score network and natural score network are optimized by K-CROSS using a regressor in the second stage. For stage one, the source modality neuroimage and the target modality neuroimage are fed into tumor branch, complex branch and structure branch.  A shared cross-modality segmentation network and a personal tumor encoder-decoder make up the tumor branch. The segmentation network is used to obtain the predicted lesion region. The tumor encoder then captures the brightness and texture features of the lesion region. As for complex branch, the input is converted into k-space by Fourier transform. The function of the private complex encoder is to identify the k-space feature shit. The source modality and the target modality share the structure branches in K-CROSS. The structure's goal is to gather information about the shared structure from both inputs. The architecture and weights of the tumor encoder, complex encoder, and structure encoder are cloned from the first stage in the second stage. The complex score network and the natural score network are optimized by K-CROSS using a regressor.
\subsubsection{Complex Branch}
A more refined U-Net architecture implemented in k-space makes up the proposed complex encoder. Specifically, each downsampling block in the encoding stage includes complex convolution $\mathbb{C}BN$ and $\mathbb{C}LeakyRELU$. The complex convolution is replaced with the complex transposed convolution for up-sampling during the decoding phase. There is a complex transposed convolution, $\mathbb{C}BN$, and $\mathbb{C}RELU$ in each upsampling block. We apply $\mathbb{C}Upsample$, complex convolution and $\mathbb{C}Tanh$ to reconstruct the images in the final layer of the decoding stage. Fig.~\ref{fig:complex_branch} shows the complex branch architecture in detail.

\subsubsection{Tumor and Structure Branch}
As a cross-modality segmentation neural network, the well-trained nnU-Net~\citep{Isensee2020nnUNetAS} is used in K-CROSS, with the weights being adjusted in the second stage of training. The modality-specific tumor encoder and decoder are private because the tumor information (the texture details and brightness) from the source modality and the target modality differ. With the exception of the operators using the normal convolution, batch norm, and Leaky RELU, the architecture details of the tumor encoder-decoder and the structure encoder-decoder are similar to those of Fig.~\ref{fig:complex_branch}. 

\subsubsection{Score Network and Quality Prediction Regressor}
We construct a two-layer MLP for quality prediction, considering that the regressor simply maps the output vectors of the triple-path decoder to labeled quality scores. The network is made up of two fully connected layers with 512-256 and 256-1 channels. The complex score network is composed of two complex fully interconnected layers. Its structure is similar to that shown in Fig.~\ref{fig:complex_op_style}. However, the operator of the complex score network substitutes MLP layers for $\mathbb{C}RELU$. The natural score network has two fully connected layers as well. The channels of the complex score network and the natural score network are 512-256 and 256-1, respectively. The regressor is trained by using the $L_1$ loss function.

\subsection{Loss Function}

\subsubsection{Frequency Loss}
\begin{figure}[ht]
	\centerline{\includegraphics[width=1\linewidth]{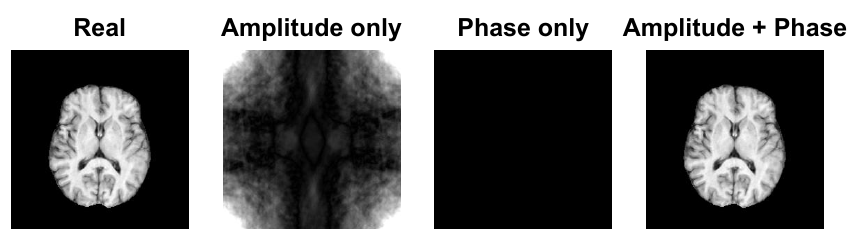}}
	\caption{Illustration of the k-space amplitude and phase of images. When only the amplitude or phase is given, the reconstructed image will lose the information of the real image.  
	}\label{fig:kspace_ample_phase}
\end{figure}

Directly measuring the distance between two complex vectors is very difficult. Alternatively, recent works are more concerned with the image amplitude. We discover that without the phase information from Fig.~\ref{fig:kspace_ample_phase}, which is impossible to reconstruct the entire neuroimage. 

\begin{figure}[ht]
	\centerline{\includegraphics[width=0.75\linewidth]{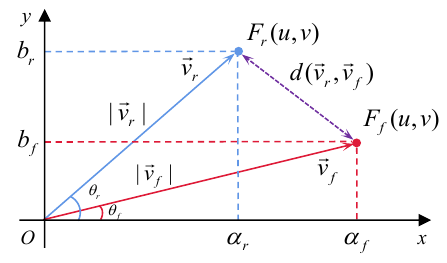}}
	\caption{Frequency loss: the distance between $\vec{v_{r}}$ and $\vec{v_{f}}$ corresponds to the distance between $F_{r}$ and $F_{f}$.
	}\label{fig:frequency_loss}
\end{figure}
Our solution is based on the focal frequency loss~\citep{Jiang2021FocalFL}, as shown in Fig.~\ref{fig:frequency_loss}. 
The hidden k-space of real MRI is $F_{r} (u, v) = a_{r} + b_{r}i$, and the corresponding k-space of synthesis MRI is $F_{f} (u, v) = a_{f} + b_{f}i$. To calculate their distance, we map $F_{r}$ and $F_{f}$ into the Euclidean space as $\vec{v_{r}}$ and $\vec{v_{f}}$. Speciﬁcally, the lengths of $\vec{v_{r}}$ and $\vec{v_{f}}$ are the amplitudes of $F_{r}$ and $F_{f}$, respectively. And the angles $\theta_r$ and $\theta_f$ correspond to the phases of $F_r$ and $F_f$, respectively. As a result, the distance between $F_r$ and $F_f$ can be converted to the distance between $\vec{v_{r}}$ and $\vec{v_{f}}$ (termed as $d(\vec{v_{r}}, \vec{v_{f}})$), which is defined as follows:
\begin{equation}
    d(F_{r}, F_{i}) = d(\vec{v_{r}}, \vec{v_{f}}) = \left\| \vec{v_{r}} - \vec{v_{f}}\right\|^{2}.
\end{equation}
The complex feature maps are extracted from each layer $l$ of the encoder in the complex U-Net. Each pixel of the complex feature maps for each layer is denoted as $m^{l} \in \mathbb{R}^{H_{l} \times W_{l} \times C_{l}}$. Finally, we compute spatial and channel averages. As a result, the frequency loss for a complex U-Net is defined as follows:
\begin{equation}\label{eq:freq_loss}
     \mathcal{L}_{freq}(m^{l}_{r}, m^{l}_{f}) = \sum_{l}\frac{1}{H_{l}W_{l}}\sum_{h, w} \left\| \vec{v_{r}} - \vec{v_{f}}\right\|^{2}.
\end{equation}

\subsubsection{Similarity Loss}
For the similarity loss $ \mathcal{L}_{simi} $, K-CROSS uses the maximum mean discrepancy (MMD) loss~\citep{Simonyan2015VeryDC} to measure it. That is, K-CROSS computes the squared population MMD between shared structure encoding of the source modality $h^{s}_{c}$ and the target modality $h^{s}_{t}$ using a biased statistic. We express this as:
\begin{equation}\label{eq:simi}
\small
    \begin{aligned}
           \mathcal{L}_{simi} = &\frac{1}{(N^{s})^{2}} \sum_{i,j=0}^{N^{s}} \kappa(h^{s}_{c_{i}}, h^{s}_{c_{j}})
            - \frac{2}{N^{s}N^{t}} \sum_{i,j=0}^{N^{s}, N^{t}} \kappa(h^{s}_{c_{i}}, h^{s}_{c_{j}})\\
           &+ \frac{1}{(N^{t})^{2}} \sum_{i,j=0}^{N^{t}} \kappa(h^{s}_{c_{i}}, h^{s}_{c_{j}}), \\
    \end{aligned}
\end{equation}
where $\kappa$ is a linear combination of multiple RBF kernels: $\kappa(x_{i}, x_{j}) = \sum_{n} \eta_{n} \mathrm{exp} \left\{ - \frac{1}{2\sigma} \left\| x_{i} - x_{j}\right\|^{2} \right\}$, where $\sigma_{n}$ is the standard deviation and $\eta_{n}$ is the weight for $n$-th RBF kernel. The similarity loss function encourages the shared structure encoder to learn the invariant structure feature irrespective of the modality.

\subsubsection{Tumor Loss} 
The tumor loss function consists of a Laplacian loss function $\mathcal{L}_{lap}$ and the LPIPS loss function $\mathcal{L}_{lpips}$~\citep{Zhang2018TheUE}. The Laplacian loss function is defined as : 
\begin{equation}
    \mathcal{L}_{lap} = \mathbb{E}\left\| L(x) - L(\hat{x})\right\|^{2}_{2}.
\end{equation}
The LPIPS loss function is defined as:
\begin{equation}\label{eq:lpips}
    \mathcal{L}_{lpips} = \sum_{k} \tau^{k} (\phi^{k}(x) - \phi^{k}(\hat{x})).
\end{equation}
So the tumor loss is described below:
\begin{equation}\label{eq:tumor}
    \mathcal{L}_{tumor} = \lambda_{lap}\mathcal{L}_{lap} + \lambda_{lpips}\mathcal{L}_{lpips}.
\end{equation}

In~\eqref{eq:lpips}, $\phi(\cdot)$ represents the feature extractor and $\tau(\cdot)$ computes the feature score from the k-th layer of the backbone architecture. As a result, the LPIPS value is the average score of all backbone layers. To compute the LPIPS loss, we used a well-trained VGG~\citep{Simonyan2015VeryDC} network. The Laplacian loss is used to identify the tumor region's high-frequency component. Due to LPIPS loss, the real tumor region and the reconstructed tumor region are more similar, which is more consistent with the radiologist's judgment.

\subsubsection{Structure Loss} 
We employ $L_{1}$ loss function to extract meaningful semantic structure features, where the structure loss function is defined as: 
\begin{equation}\label{eq:stru}
    \mathcal{L}_{stru} = ||x - \hat{x}||_1.
\end{equation}

\subsubsection{Inconsistency Loss} We adopt the MSE loss function to optimize the weights of complex score network $n_{c}$ and natural score network $n_{nat}$, where the inconsistency loss is defined as:
\begin{equation}\label{eq:inconsistency_loss}
    \mathcal{L}_{inc} =  ||\eta_{total} - \eta_{ra}||_1,
\end{equation}
where the score of K-CROSS $\eta_{total}$ is aligned with the scale of the radiologist's rating score $\eta_{ra}$ via our proposed ranking algorithm. The details of the ranking algorithm can be found in Algorithm~\ref{alg:ranking}.

\subsubsection{Total Loss}
For the first stage, the loss function is described below:
\begin{equation}
    \mathcal{L}_{first} = \lambda_{1}\mathcal{L}_{tumor} + \lambda_{2}\mathcal{L}_{stru} + \lambda_{3}\mathcal{L}_{freq} + \lambda_{4}\mathcal{L}_{sim}.
\end{equation}

In the second stage, we optimize the parameters of the complex score network and the natural score network via
\begin{equation}
\mathcal{L}_{second} = \mathcal{L}_{inc}.
\end{equation}
In this work, all weights of $\lambda$ are set to 1.

\begin{table}[t]
    \centering
    \caption{
    Description for notations used in our algorithms.
	}
     \renewcommand{\arraystretch}{1.2}
    \begin{tabular}{l|l}
    \hline
    \rowcolor{blue!5}\textbf{Notation} & \textbf{Description} \\\hline
    $s $ & Source neuroimage \\
    $t$ & Paired target neuroimage \\
    $\hat{t}$    &  Synthesized target modality neuroimage \\
    $\Omega_{ref}$ & Reference images \\
    $\Omega_{syn}$ & Synthesised images \\
    $\varphi_{E}$ & Tumor encoder \\
    $\varphi_{D}$ & Source tumor decoder \\
    $\psi_{E}$    & Source complex encoder \\
    $\psi_{D}$    & Complex decoder \\
    $\xi_{E}$     & Shared structure encoder \\
    $\xi_{D}$     & Shared structure decoder\\
    $\nu$         & Segmentation network \\
    $FT$          & Fourier transform \\
    $l$           & Lesion mask \\
    $\zeta$       & K-space function \\
    $z$           & Structural feature  \\
    $\theta^{n}_{X}$    & The parameters of $X$ trained in stage $n$ \\
    $n_{c}$      &  Complex score network \\
    $n_{nat}$    &  Natural score network \\
    $\eta_{ra}$  & Radiologist rating score  \\
    $\eta_{total}$ & K-CROSS score \\
    \hline
    \end{tabular}
\label{tab:notation}
\end{table}

\begin{algorithm}[t]
\small
\caption{K-CROSS training stage 1}\label{alg:kcross_training_1}
    \SetKwInOut{KwIn}{Input}
    \SetKwInOut{KwOut}{Output}
    \KwIn{Source neuroimage $s$, paired target neuroimage $t$, tumor encoder $\varphi_{E}$, source tumor decoder $\varphi_{D}$, source complex encoder $\psi_{E}$, complex decoder $\psi_{D}$, shared structure encoder $\xi_{E}$, shared structure decoder $\xi_{D}$, segmentation network $\nu$, Fourier transform $FT$, lesion mask $l$, k-space $\zeta$, structural feature $z$.}
    \KwOut{The parameters of tumor encoder $\theta^{1}_{\varphi_{E}}$, tumor decoder $\theta_{\varphi_{D}}$, complex encoder $\theta^{1}_{\psi_{E}}$, complex decoder $\theta_{\psi_{D}}$, structure encoder $\theta^{1}_{\xi_{E}}$ and structure decoder $\theta_{\xi_{D}}$. }
    $l_{t} \leftarrow \nu(t)$ \textcolor{gray}{// generate lesion mask of target neuroimage }\\
    $\zeta_{t} \leftarrow FT(t)$ \textcolor{gray}{// convert k-space of target neuroimage  } \\
    $z_{s} \leftarrow \xi_{E}(s)$ \textcolor{gray}{// compute structure feature of source neuroimage} \\
    $z_{t} \leftarrow \xi_{E}(t)$ \textcolor{gray}{// convert structure feature of target neuroimage  } \\
    
    \While{not converged}
    {
        Update $\theta^{1}_{\psi_{E}}$ and $\theta^{1}_{\psi_{D}}$ with $\kappa_{t}$, $\mathcal{L}_{freq}$ in~\eqref{eq:freq_loss} \\
        Update $\theta^{1}_{\varphi_{E}}$ and $\theta^{1}_{\varphi_{D}}$ with $l_{t}$, $\mathcal{L}_{tumor}$ in~\eqref{eq:tumor}\\
        Update $\theta^{1}_{\xi_{E}}$ and $\theta_{\xi_{D}}$ with $z_{s}$, $z_{t}$, $\mathcal{L}_{simi}$ in~\eqref{eq:simi} and $\mathcal{L}_{stru}$ in~\eqref{eq:stru}\\
       
    }
\end{algorithm}

\begin{algorithm}[ht]
\small
\caption{K-CROSS training stage 2}\label{alg:kcross_training_2}
    \SetKwInOut{KwIn}{Input}
    \SetKwInOut{KwOut}{Output}
    \KwIn{Synthesis target modality neuroimage $\hat{t}$, tumor encoder and its parameters from stage 1: $\varphi_{E}$ and $\theta^{1}_{\varphi_{E}}$, complex encoder and its parameters from stage 1: $\psi_{E}$ and $\theta^{1}_{\psi_{E}}$, structure encoder and its parameters from stage 1: $\xi_{E}$ and $\theta^{1}_{\xi_{E}}$, complex score network $n_{c}$, natural score network $n_{nat}$. }
    \KwOut{The parameters of tumor encoder $\theta^{2}_{\varphi_{E}}$, complex encoder $\theta^{2}_{\psi_{E}}$, structure encoder $\theta^{2}_{\xi_{E}}$, complex score network $\theta_{n_{c}}$, natural score network $\theta_{n_{nat}}$  }

    $\zeta_{\hat{t}} \leftarrow FT(\hat{t})$ \textcolor{gray}{// convert k-space of synthesis target neuroimage } \\
    $l_{\hat{t}} \leftarrow \nu(\hat{t})$ \textcolor{gray}{// generate lesion mask of 
    synthesis target neuroimage }\\
    $f_{complex} \leftarrow \psi_{E} (\zeta_{\hat{t}})$ \textcolor{gray}{// compute the complex feature } \\
    $f_{tumor} \leftarrow \varphi_{E} (l_{\hat{t}})$ \textcolor{gray}{// compute the tumor feature } \\
    $f_{stru} \leftarrow \xi_{E} (\hat{t})$ \textcolor{gray}{// compute the structure feature} \\
    $\eta_{complex} \leftarrow n_{c}(f_{complex})$ \textcolor{gray}{// compute the complex score} \\
    $\eta_{nat} \leftarrow n_{nat} (f_{tumor} + f_{stru})$ \textcolor{gray}{// compute the nature score} \\
    $\eta_{total} \leftarrow \eta_{nat} + \eta_{complex}$ \\
    Ranking all $\eta_{total}$ and get in line with $\eta_{ra}$ \\
    \While{not converged}
    {
        Update $\theta_{n_{c}}$ and $\theta_{n_{nat}}$ with $\eta_{total}$, $\eta_{ra}$ and $\mathcal{L}_{inc}$ in~\eqref{eq:inconsistency_loss}
    }
    
\end{algorithm}

\begin{algorithm}[ht]
\small
\caption{K-CROSS inference}\label{alg:kcross_infer_1}
    \SetKwInOut{KwIn}{Input}
    \SetKwInOut{KwOut}{Output}
    \KwIn{Synthesis target modality neuroimage $\hat{t}$, tumor encoder and its parameters from stage 1: $\varphi_{E}$ and $\theta^{1}_{\varphi_{E}}$, complex encoder and its parameters from stage 1: $\psi_{E}$ and $\theta^{1}_{\psi_{E}}$, structure encoder and its parameters from stage 1: $\xi_{E}$ and $\theta^{1}_{\xi_{E}}$, complex score network and its parameters from stage 2: $n_{c}$ and $\theta_{n_{c}}$, natural score network and its parameters from stage 2: $n_{nat}$ and $\theta_{n_{nat}}$, Fourier transform $FT$, segmentation network $\nu$.  }
    \KwOut{K-CROSS score $\eta_{total}$.  }
    $\zeta_{\hat{t}} \leftarrow FT(\hat{t})$ \textcolor{gray}{// convert k-space of synthesis target neuroimage } \\
    $l_{\hat{t}} \leftarrow \nu(\hat{t})$ \textcolor{gray}{// generate lesion mask of synthesis target neuroimage }\\
    $f_{complex} \leftarrow \psi_{E} (\zeta_{\hat{t}})$ \textcolor{gray}{// compute the complex feature  } \\
    $f_{stru} \leftarrow \xi_{E} (\hat{t})$ \textcolor{gray}{// compute the structure feature} \\
    $\eta_{complex} \leftarrow n_{c}(f_{complex})$ \textcolor{gray}{// compute the complex score} \\
    \eIf{$\hat{t} \in $ healthy person}
    {
        $\eta_{nat} \leftarrow f_{stru}$
    }
    {
       $l_{\hat{t}} \leftarrow \nu(\hat{t})$ \textcolor{gray}{// generate lesion mask of synthesis target neuroimage }\\
       $f_{tumor} \leftarrow \varphi_{E} (l_{\hat{t}})$ \textcolor{gray}{// compute the tumor feature} \\
       $\eta_{nat} \leftarrow n_{nat} (f_{tumor} + f_{stru})$  \\
    }
    $\eta_{total} \leftarrow \eta_{nat} + \eta_{complex}$ \\
    Ranking all $\eta_{total}$ and get in line with $\eta_{ra}$
\end{algorithm}

\begin{algorithm}[t]
\small  
	\caption{Calculate inconsistency by ranking}\label{alg:ranking}
	\textbf{Input:}{ reference images $\Omega_{ref}$, synthesised images $ \Omega_{syn} $ }\\
	\textbf{Initialize:} start = 0, pairwise $\leftarrow \phi$, uniform\_result $\leftarrow \phi$ \\
	\textcolor{gray}{// count elements in levels}\\
	 recounted = Collection.Counter($\Omega_{ref}$) \\
	 \textcolor{gray}{// obtain the sorted level of reference images} \\
	 level = sorted(recounted)\\
	 \textcolor{gray}{// obtain the indices of synthesised images} \\
	index = numpy.argsort( $ \Omega_{syn} $) \\
	\textcolor{gray}{// obtain the number of images $\Omega_{ref}$ in each level} \\
	\For{$l$ in level}{
	   end = start + recounted[$l$] \\
	   pairwise.append([start, end])\\
	   start = end\\
	}
    \textcolor{gray}{// obtain uniformed results $ \Omega_{syn} $}\\
    \For{v in index}{
        \For{i, p in enumerate(pairwise)}{
            \If{ v in range(p[0], p[1])}{  
                uniform\_result.append(i / len(level)) \\
                break 
            }
        }
	} 
    \textcolor{gray}{// calculate consistency}\\
    \textbf{Return:} $ ||\Omega_{ref} - \Omega_{syn}||_1$.mean()
\end{algorithm}

\subsection{Algorithms}\label{sec:alg}
The two stages of training K-CROSS are depicted in Fig~\ref{fig:K-CROSS}. The details of two-stage training algorithms and the inference algorithm are described in Algorithm~\ref{alg:kcross_training_1}, Algorithm~\ref{alg:kcross_training_2} and Algorithm~\ref{alg:kcross_infer_1}, respectively. For clarity, Table~\ref{tab:notation} provides notation descriptions that occurred in our algorithms.

\section{NIRPS Dataset and Radiologist Score}\label{sec:nirps-dataset}

To comprehensively evaluate the synthesis performance, we construct a large-scale multi-modal neuroimaging perceptual similarity (NIRPS) dataset with 6,000 radiologist judgments. NIRPS dataset is composed of three subsets generated by CycleGAN~\citep{zhu2017unpaired}, MUNIT \cite{huang2018multimodal} and UNIT \cite{DBLP:conf/nips/LiuBK17}. Each set contains 800 images generated by IXI and 1,200 images generated by BraTS. The IXI dataset includes two modalities, PD and T2, while the BraTS dataset includes three modalities, T1, T2, and FLAIR. In both the IXI and BraTS datasets, we randomly select 10 slices for training and collect the training results after each epoch of the model trained over 40 epochs.

\textbf{IXI}~\citep{Aljabar2011ACM} collects nearly 600 MR images from normal and healthy subjects in three hospitals. The MR image acquisition protocol for each subject includes T1, T2, PD-weighted images (PD), MRA images, and Diffusion-weighted images. In this paper, we only use T1 (581 cases), T2 (578 cases) and PD (578 cases) data to conduct our experiments, and select the paired data with the same ID from the three modes. The image has a non-uniform length on the z-axis with the size of $ 256 \time 256 $ on the x-axis and y-axis. The IXI dataset is not divided into a training set and a test set. Therefore, we randomly split the whole data as the training set (80\%) and the test set (20\%).

\textbf{BraTS2021}~\citep{Siegel2019CancerS2,Bakas2017BrainlesionGM} is designed for the analysis and diagnosis of brain diseases. The dataset of multi-institutional and pre-operative MRI sequences is made publicly available and includes both training data (1251 cases) and validation data (219 cases). Each 3D volume is 155$\times$240$\times$240 in size and is imaged by four sequences: T1, T2, T1ce, and FLAIR.

\textbf{Training Data Processing} To ensure the validity and diversity of the data, we remove their skulls for each slice, by splitting the three-dimensional volume and choosing slices ranging from 50 to 80 on the z-axis. All images are cropped to  $ 256 \time 256 $ pixels in size. During the training stage, we choose a total of 10k images from the IXI and BraTS2021 datasets.

 \begin{figure}[ht]
	\centerline{\includegraphics[width=0.75\linewidth]{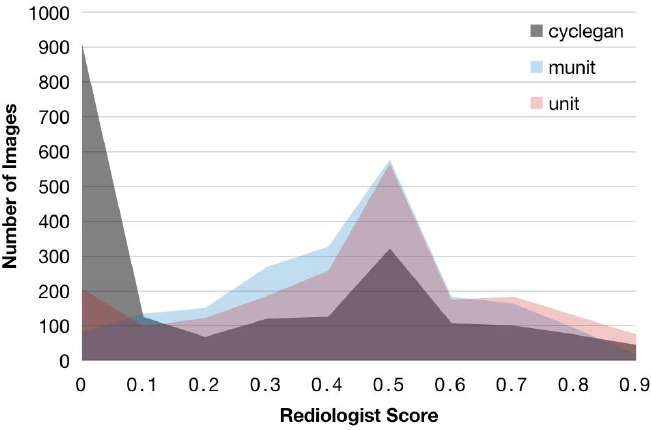}}
	\caption{
	Statistical results of NIRPS.
	}\label{fig:nirps_dataset}
\end{figure}

\subsubsection{Radiologist Score} The NIRPS dataset contains radiologist scores (RS) resulting from manual annotation for each image. It is worth noting that the radiologist score $RS$ includes 10 levels, i.e., $ RS \in [0, 0.1, 0.2, .., 0.9]$. The higher RS value indicates better-synthesized neuroimage quality. Radiologists give scores according to the level of diagnosis and therapy using the synthesized neuroimage. Fig.~\ref{fig:nirps_dataset} gives the distribution result of $RS$. We can see that synthesized performance varies among the three models and the average $RS$ is in the middle. 

\subsubsection{How Radiologists Assess?} 
\begin{figure}[ht]
	\centerline{\includegraphics[width=0.75\linewidth]{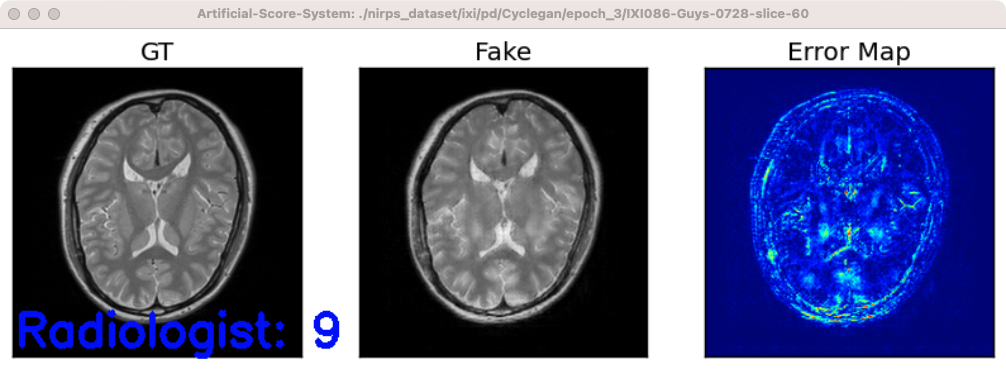}}
	\caption{
	Labelling tools in our artificial score system. The first column denotes the ground-truth T2 neuroimage. The second column denotes the synthesized T2 neuroimage. The third column is the error map between the ground truth neuroimage and its synthesized one.               
	}\label{fig:nirps_labeling_tool}
\end{figure}

We prepare the real paired modalities neuroimage dataset $M$ in advance. $M$ consists of source modalities $M_{s}$ and target modalities $ M_{t} $. We generate the synthesized target modality neuroimages $\hat{M}_{t}$ via feeding $M_{s}$ into the generative model, i.e., CycleGAN, MUNIT and UNIT in NIPRS. Then radiologist gives the score for $\hat{M}_{t}$ according to the comparison with $\hat{M}_{t}$. For instance, we have paired ground-truth modality datasets, T1 and T2. As shown in Fig.~\ref{fig:nirps_labeling_tool}, we synthesized the fake T2 by feeding T1 into the MUNIT model. The radiologists make direct comparisons between fake T2 and real T2 and give their score for the synthesised quality of T2.

\subsubsection{How Radiologists Combine Their Evaluations?} We hire 10 radiologists to evaluate the quality of each synthesized neuroimage. We remove the highest score and the lowest score from all radiologists. The final score is then averaged by the rest score from 8 radiologists. 

\section{Experiment and Ablation Study}\label{sec:experiment}

\begin{table}
    \caption{
    Performance comparison of various IQA methods on NIRPS. NIRPS includes three subsets generated by CycleGAN, MUNIT, and UNIT. Specifically, each set consists of 800 images for IXI and 1,200 images for BraTS. Here, IXI and BraTS training sets are used to train our model. The inconsistency is computed, where $\downarrow$ is better. The best result is highlighted in red.
	}
 \renewcommand{\arraystretch}{1.3}
\resizebox{\linewidth}{!}{
\begin{tabular}{l|cc|cc|cc}
\hline
\rowcolor{blue!5}\textbf{Metric} & \multicolumn{2}{c|}{\textbf{CycleGAN}} &\multicolumn{2}{c|}{\textbf{MUNIT}} & \multicolumn{2}{c}{\textbf{UNIT}}\\\hline
        \rowcolor{blue!5}Dasteset & IXI        & BraTS      & IXI       & BraTS    & IXI      & BraTS   \\ \hline
        MAE~\citep{Chai2014RootMS}     & 0.3127     & 0.3335     & 0.1807    & 0.2368   & 0.2462   & 0.2462  \\
        PSNR~\citep{HuynhThu2008ScopeOV}    & 0.3387     & 0.2060     & 0.1832    & 0.2515   & 0.2387   & 0.2585  \\
        SSIM~\citep{Wang2004ImageQA}    & 0.3333     & 0.1998     & 0.1957    & 0.2437   & 0.2523   & 0.2645  \\
        MS-SSIM~\citep{Wang2003MultiscaleSS} & 0.3245     & 0.1973     & 0.1934    & 0.2425   & 0.2487   & 0.2521  \\
        NLPD~\citep{Laparra2016PerceptualIQ}    & 0.3128     & 0.1961     & 0.1921    & 0.2408   & 0.2466   & 0.2445  \\
        GMSD~\citep{Xue2014GradientMS}    & 0.3069     & 0.1952     & 0.1910    & 0.2395   & 0.2421   & 0.2432  \\
        DeepIQA~\citep{Bosse2018DeepNN} & 0.3052     & 0.1943     & 0.1827    & 0.2380   & 0.2402   & 0.2410  \\
        LPIP~\citep{Zhang2018TheUE}    & 0.3023     & 0.1921     & 0.1786    & 0.2285   & 0.2398   & 0.2327  \\
        DIST~\citep{Ding2022ImageQA}    & 0.3012     & 0.1901     & 0.1725    & 0.2291   & 0.2387   & 0.2230  \\
\hline
\rowcolor{black!5} \textbf{K-CROSS} & \textcolor{red}{0.2878}     & \textcolor{red}{0.1851}     & \textcolor{red}{0.1650}    & \textcolor{red}{0.2246}   & \textcolor{red}{0.2123}   & \textcolor{red}{0.2147}\\
\hline
\end{tabular}}

\label{tab:overall_performance}
\end{table}

\subsection{K-CROSS vs Other Metrics}
Table~\ref{tab:overall_performance} illustrates the inconsistency between metrics and human evaluations of several datasets and generative models, with the highest performance shown in red. The calculation method for inconsistency value is given in Algorithm~\ref{alg:ranking}. We evaluate K-CROSS on datasets created by CycleGAN, MUNIT, and UNIT. The first column indicates various IQA methods. The second column indicates which datasets were used to train the K-CROSS model, including IXI or BraTS. From Table~\ref{tab:overall_performance}, our proposed K-CROSS is more compatible with the assessments of radiologists. Note that the IXI dataset is a healthy person dataset. There is no lesion for each neuroimage. So K-CROSS only use the tumor branch and complex branch to assess the quality of neuroimage. The details are described in Section~\ref{sec:healthy_person}.

\subsection{K-Space Importance}
Table~\ref{tab:k-importance} records the ablation study of individual branches (the complex branch, tumor branch and structure branch) on various datasets. For example, when we conduct the ablation study of the complex branch, K-CROSS removes the tumor branch and the structure branch in the inference phase. In other words, K-CROSS only obtains $\eta_{complex}$ score. It applies the same setting for the other branches. It can be clearly observed that the complex branch obtains the highest score among the three branches. It strongly indicates the importance of k-space, which reflects the inherent properties of magnetic resonance imaging principles. The second best is the tumor branch. It also verifies the effectiveness of the tumor branch for the lesion disease dataset.

\begin{table}[ht]
\centering
\caption{Effect of k-space importance (complex branch) and tumor branch. The inconsistency is computed, where $\downarrow$ is better. The best result is in red and the second best result is in blue.}
\renewcommand{\arraystretch}{1.3}
\resizebox{0.85\linewidth}{!}{
\begin{tabular}{l|cccc}
\hline
\rowcolor{blue!5}  \textbf{Model} & \makecell[c]{\textbf{Structure Branch }\\ $\mathcal{L}_{stru}+\mathcal{L}_{simi}$} & \makecell[c]{\textbf{Complex Branch} \\ $\mathcal{L}_{freq}$ } & \makecell[c]{\textbf{Tumor Branch}\\ $\mathcal{L}_{tumor}$} & \textbf{All}\\  
\hline

\multicolumn{5}{c}{\textit{BraTS T1$\leftrightarrow$FLAIR}} \\
\hline
CycleGAN              & 0.1970    & \textcolor{blue}{0.1911}    & 0.1920            & \textcolor{red}{0.1880}                     \\
MUNIT       & 0.2348                     & \textcolor{blue}{0.2273}         & 0.2338          &  \textcolor{red}{0.2250}         \\
UNIT           & 0.2333   & \textcolor{blue}{0.2210}       &  0.2248   &  \textcolor{red}{0.2102}        \\
\hline

\multicolumn{5}{c}{\textit{BraTS T2$\leftrightarrow$FLAIR}} \\
\hline
CycleGAN              & 0.1983    & \textcolor{blue}{0.1889}                & 0.1953            &  \textcolor{red}{0.1853}                     \\
MUNIT       & 0.2365          & \textcolor{blue}{0.2272}                   & 0.2348          &  \textcolor{red}{0.2250}         \\
UNIT           & 0.2305   & \textcolor{blue}{0.2250}       & 0.2297    &  \textcolor{red}{0.2155}        \\
\hline
\multicolumn{5}{c}{\textit{BraTS T1$\leftrightarrow$T2}} \\
\hline
CycleGAN              & 0.1839      & \textcolor{blue}{0.1836}               & 0.1953            &  \textcolor{red}{0.1819}                     \\
MUNIT       & 0.2332          & \textcolor{blue}{0.2258}                & 0.2308          &  \textcolor{red}{0.2239}         \\
UNIT           & 0.2595   & \textcolor{blue}{0.2290}       & 0.2298    &  \textcolor{red}{0.2183}        \\
\hline

\end{tabular}
}\label{tab:k-importance}
\end{table}

\subsection{Metrics for Healthy Person}\label{sec:healthy_person}
Table~\ref{tab:healthy_person} shows K-CROSS performance that surpasses the mainstream IQA methods on the IXI healthy-person dataset. As for assessing the synthesised neuroimage of healthy persons, K-CROSS removes the tumor branch in the inference phase. Because there are no lesions on healthy person datasets. It means that K-CROSS only combines $\eta_{complex}$ and the score of the structure encoder as the final score. From Table~\ref{tab:healthy_person}, it can be obviously observed that K-CROSS complex branch score $\eta_{complex}$ (blue value) has surpassed the other IQA methods, which identify the importance of k-space for MRI of healthy persons. Thus, K-CROSS still be able to serve as the metric for the synthesized quality of healthy person's neuroimage.

\begin{table}[t]
\centering
\caption{Metric performance for healthy persons on IXI. The inconsistency is computed, where $\downarrow$ is better. The best result is in red and the second best result is in blue.}
\renewcommand{\arraystretch}{1.3}
\resizebox{0.85\linewidth}{!}{
\begin{tabular}{l|c|c|c}
\hline
\rowcolor{blue!5}\textbf{Metric} & \textbf{CycleGAN} &\textbf{MUNIT} & \textbf{UNIT}\\\hline
    MAE~\citep{Chai2014RootMS}    &0.3160	&0.1809	  &0.2442                   \\
    PSNR~\citep{HuynhThu2008ScopeOV}   &0.3388	&0.1835	&0.2419                    \\
    SSIM~\citep{Wang2004ImageQA}   &0.3352	&0.1962	&0.2423                   \\
    MS-SSIM~\citep{Wang2003MultiscaleSS}                     &0.3238		&0.1945		&0.2418                    \\
    NLPD~\citep{Laparra2016PerceptualIQ}   &0.3135	&0.1918		&0.2409                    \\
    GMSD~\citep{Xue2014GradientMS}  & 0.3090		&0.1908		&0.2398                    \\
    DeepIQA~\citep{Bosse2018DeepNN}                       &0.3043	&0.1856	 	&0.2381                   \\
    LPIP~\citep{Zhang2018TheUE}   &0.3022	&0.1810	             	 &0.2374                    \\
    DIST~\citep{Ding2022ImageQA}  &0.3056	&0.1805    	&0.2356                   \\ \hline
\rowcolor{black!5} \textbf{K-CROSS ($\mathcal{L}_{stru}$)}            & 0.3078         & 0.1790         & 0.2318                            \\
\rowcolor{black!5} \textbf{K-CROSS($\mathcal{L}_{comp}$)}                     & \textcolor{blue}{0.3045}         &  \textcolor{blue}{0.1618}         &  \textcolor{blue}{0.2233}                             \\
\rowcolor{black!5} \textbf{K-CROSS ($\mathcal{L}_{stru}$+$\mathcal{L}_{comp}$)} &  \textcolor{red}{0.2876}         &  \textcolor{red}{0.1610}         &  \textcolor{red}{0.2150}                             \\ \hline
\end{tabular}
}\label{tab:healthy_person}
\end{table}

\begin{figure*}[t]
	\centerline{\includegraphics[width=1\linewidth]{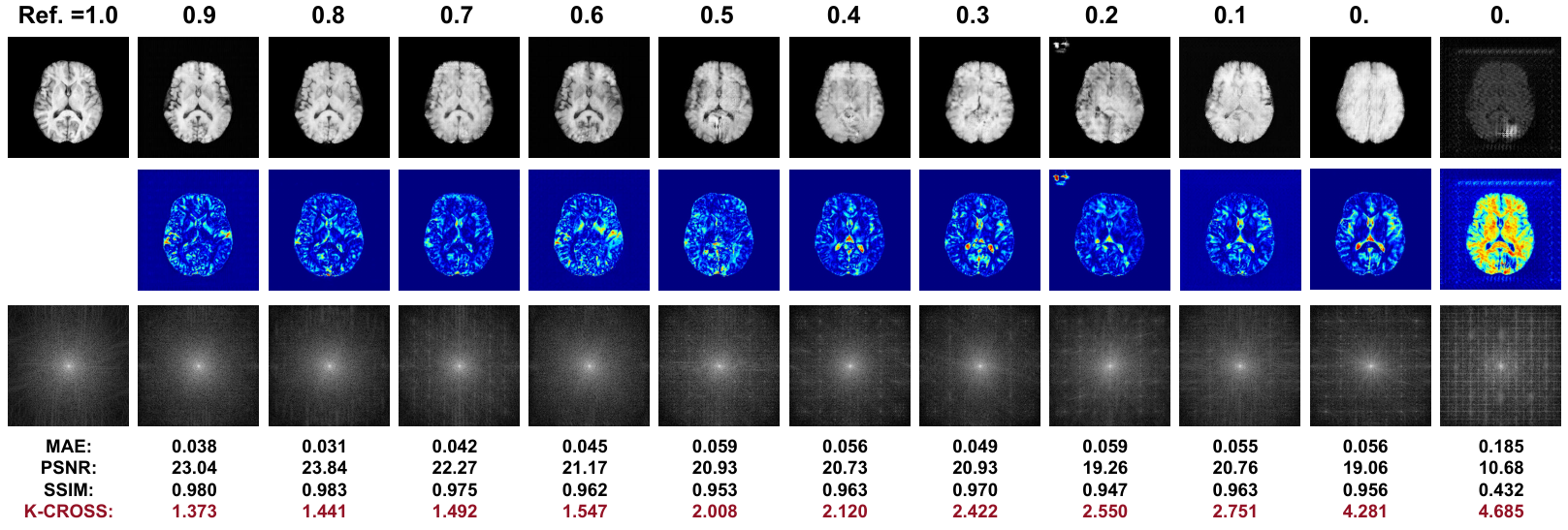}}
	\caption{
	Visualization of the NIRPS dataset. The number above each synthesized neuroimage is the radiologist's score. The blue map denotes the error map between the synthesis neuroimage data and the reference image. The third row represents the k-space map. A higher value of MAE, PSNR, SSIM and radiologist' score indicates higher performance of the synthesis quality. A lower value of K-CROSS indicates a higher performance of the synthesis quality. It's easily observed that K-CROSS is consistent with the order of the radiologist's score when the synthesis quality deteriorates but PSNR, SSIM and MAE are not.
	}\label{fig:vis_nirps_dataset}
\end{figure*}

\subsection{Segmentation Network Effect}
Table~\ref{tab:segmentation_backbone} shows K-CROSS remains stable performance even using different state-of-the-art medical segmentation models. Note that the parameters of the pre-trained segmentation network are frozen during the training phase. The first column denotes the segmentation method. We calculate the variance score of the K-CROSS value for CycleGAN, MUNIT and UNIT by using different segmentation backbone models. We find that the variance of K-CROSS performance is tiny (0.2\%, 0.3\%, and 0.2\%). Hence, the performance of K-CROSS is not affected by the segmentation model. 

\begin{table}[ht]
\caption{The effect of segmentation network for K-CROSS value for various generative models of NIRPS.}
\centering
\renewcommand{\arraystretch}{1.3}
\resizebox{1\linewidth}{!}{
\begin{tabular}{l|c|c|c}
\hline
\rowcolor{blue!5} \textbf{Segmentation Model} & \textbf{CycleGAN} & \textbf{MUNIT} & \textbf{UNIT} \\ \hline
nnUnet~\citep{Isensee2020nnUNetAS}               & 0.1872            & 0.2246         & 0.2160                               \\
AttnUnet~\citep{Schlemper2019AttentionGN}      & 0.1873            & 0.2268         & 0.2157                               \\
SETR~\citep{Zheng2021RethinkingSS}            & 0.1852            & 0.2254         & 0.2162                               \\
CoTr~\citep{Xie2021CoTrEB}          & 0.1856            & 0.2244         & 0.2154                               \\
TransUNet~\citep{Chen2021TransUNetTM}       & 0.1862            & 0.2247         & 0.2162                               \\
SwinUNet~\citep{Cao2021SwinUnetUP}       & 0.1845            & 0.2252         & 0.2158                               \\ \hline
\end{tabular}
}\label{tab:segmentation_backbone}
\end{table}

\begin{table}[t]
\centering
\caption{K-CROSS on NIRPS (CycleGAN). The inconsistency is computed, where $\downarrow$ is better. The best result is in red and the second best result is in blue.}
\renewcommand{\arraystretch}{1.3}
\resizebox{0.85\linewidth}{!}{
\begin{tabular}{l|c|c|c}
\hline
 \rowcolor{blue!5} \textbf{Training Set (BraTS)} &  \textbf{T1-FLAIR} & \textbf{T2-FLAIR} & \textbf{T1-T2}\\
\rowcolor{blue!5} \textbf{Test Set (IXI)} & \textbf{PD-T2} & \textbf{PD-T2} & \textbf{PD-T2}\\\hline

MAE~\citep{Chai2014RootMS}    & 0.3208                                & 0.3321                                & 0.3228                    \\
PSNR~\citep{HuynhThu2008ScopeOV}  & 0.3190  & 0.3270                                & 0.3261                    \\
SSIM~\citep{Wang2004ImageQA}   & 0.3231       & 0.3231                                & 0.3212                    \\
MS-SSIM~\citep{Wang2003MultiscaleSS}      & 0.3189       & 0.3180                                & 0.3128                    \\
NLPD~\citep{Laparra2016PerceptualIQ}  & 0.3176     & 0.3163                                & 0.3090                    \\
GMSD~\citep{Xue2014GradientMS}   & 0.3067        & 0.3154                                & 0.3067                    \\
DeepIQA~\citep{Bosse2018DeepNN}         & 0.3058            & 0.3124                                & 0.3045                    \\
LPIP~\citep{Zhang2018TheUE} & 0.3033          & 0.3056                                & 0.3039                    \\
DIST~\citep{Ding2022ImageQA}& 0.3021           & 0.3032                                & 0.3006                    \\ \hline
\rowcolor{black!5} \textbf{K-CROSS ($\mathcal{L}_{stru}$)}     &  0.3018    & 0.3013   &   0.2990 \\
\rowcolor{black!5} \textbf{K-CROSS ($\mathcal{L}_{freq}$)}                     &  \textcolor{blue}{0.3015}         &  \textcolor{blue}{0.2997}         &  \textcolor{blue}{0.2953}                             \\
\rowcolor{black!5} \textbf{K-CROSS} ($\mathcal{L}_{stru}$+$\mathcal{L}_{freq}$) &  \textcolor{red}{0.2893}         &  \textcolor{red}{0.2950}         &  \textcolor{red}{0.2890}                             \\ \hline
\end{tabular}
}\label{tab:domain_gap_cyclegan}
\end{table}

\begin{table}[htb]
\caption{K-CROSS on the NIRPS dataset (MUNIT). The inconsistency is computed, where $\downarrow$ is better. The best result is in red and the second best result is in blue.}
\renewcommand{\arraystretch}{1.3}
\resizebox{\linewidth}{!}{
\begin{tabular}{l|c|c|c}
\hline
\rowcolor{blue!5} \textbf{Training Set} &  \textbf{BraTS (T1-FLAIR)} & \textbf{BraTS (T2-FLAIR) }& \textbf{BraTS(T1-T2)}\\
\rowcolor{blue!5} \textbf{Test Set} & \textbf{IXI (PD-T2) }& \textbf{IXI (PD-T2)} & \textbf{IXI (PD-T2)}\\\hline

MAE~\citep{Chai2014RootMS}    & 0.1872        & 0.1865        & 0.1842                    \\
PSNR~\citep{HuynhThu2008ScopeOV}  & 0.1853        & 0.1873        & 0.1853                    \\
SSIM~\citep{Wang2004ImageQA}   & 0.1967        & 0.1986        & 0.1982                    \\
MS-SSIM~\citep{Wang2003MultiscaleSS}        & 0.1932        & 0.1952        & 0.1923                    \\
NLPD~\citep{Laparra2016PerceptualIQ}   & 0.1945        & 0.1943       & 0.1921                    \\
GMSD~\citep{Xue2014GradientMS}   & 0.1920        & 0.1934        & 0.1932                    \\
DeepIQA~\citep{Bosse2018DeepNN}        & 0.1835        & 0.1845        & 0.1856                    \\
LPIP~\citep{Zhang2018TheUE}  & 0.1783        & 0.1754        & 0.1784                    \\
DIST~\citep{Ding2022ImageQA}   & 0.1742        & 0.1731        & 0.1786                    \\ \hline

\rowcolor{black!5}  \textbf{K-CROSS ($\mathcal{L}_{stru}$)}     &  0.1713    & 0.1720   &    0.1790 \\
\rowcolor{black!5}  \textbf{K-CROSS ($\mathcal{L}_{freq}$)}                     &  \textcolor{blue}{0.1703}         &  \textcolor{blue}{0.1709}         &  \textcolor{blue}{0.1692}                             \\
\rowcolor{black!5}  \textbf{K-CROSS} ($\mathcal{L}_{stru}$+$\mathcal{L}_{freq}$) &  \textcolor{red}{0.1688}         &  \textcolor{red}{0.1693}         &  \textcolor{red}{0.1617}        \\ \hline

\end{tabular}
}\label{tab:domain-gap-munit}
\end{table}

\begin{table}[!h]
\caption{K-CROSS on the NIRPS dataset (UNIT). The inconsistency is computed, where $\downarrow$ is better. The best result is in red and the second best result is in blue.}
\renewcommand{\arraystretch}{1.3}
\resizebox{\linewidth}{!}{
\begin{tabular}{l|c|c|c}
\hline
\rowcolor{blue!5} \textbf{Training Set} &  \textbf{BraTS (T1-FLAIR)} & \textbf{BraTS (T2-FLAIR) }& \textbf{BraTS (T1-T2)}\\
\rowcolor{blue!5} \textbf{Test Set} & \textbf{IXI (PD-T2) }& \textbf{IXI (PD-T2)} & \textbf{IXI (PD-T2)}\\\hline

MAE~\citep{Chai2014RootMS}    & 0.2468       & 0.2445        & 0.2480                   \\
PSNR~\citep{HuynhThu2008ScopeOV}   & 0.2421        & 0.2410       & 0.2410                    \\
SSIM~\citep{Wang2004ImageQA}  & 0.2545          & 0.2505        & 0.2470                    \\
MS-SSIM~\citep{Wang2003MultiscaleSS}      & 0.2486        & 0.2474        & 0.2463                    \\
NLPD~\citep{Laparra2016PerceptualIQ}   & 0.2481        & 0.2461       & 0.2450                    \\
GMSD~\citep{Xue2014GradientMS}   & 0.2410        & 0.2388        & 0.2478                    \\
DeepIQA~\citep{Bosse2018DeepNN}        & 0.2398        & 0.2376        & 0.2391                   \\
LPIP~\citep{Zhang2018TheUE}   & 0.2352        & 0.2341        & 0.2384                    \\
DIST~\citep{Ding2022ImageQA}   & 0.2348        & 0.2337        & 0.2386                    \\ \hline

\rowcolor{black!5} \textbf{K-CROSS ($\mathcal{L}_{stru}$)}     &  0.2323    & 0.2320   &    0.2250 \\
\rowcolor{black!5} \textbf{K-CROSS ($\mathcal{L}_{freq}$)}                     &  \textcolor{blue}{0.2190}         &  \textcolor{blue}{0.2195}         &  \textcolor{blue}{0.2250}                             \\
\rowcolor{black!5} \textbf{K-CROSS} ($\mathcal{L}_{stru}$+$\mathcal{L}_{freq}$) &  \textcolor{red}{0.2188}         &  \textcolor{red}{0.2188}         &  \textcolor{red}{0.2175}        \\ \hline

\end{tabular}
}\label{tab:domain-gap-unit}
\end{table}

\subsection{General Metric? Overcoming Domain Gap}
The purpose of this paper is to demonstrate that K-CROSS is capable of serving as the standard measure for MRI datasets. We conduct extensive experiments and the results are given in Table~\ref{tab:domain_gap_cyclegan}, Table~\ref{tab:domain-gap-munit}, and Table~\ref{tab:domain-gap-unit} to demonstrate that K-CROSS is not affected by dataset domain gap and the generative model. The training dataset is the BraTS dataset, and the test dataset is IXI. As described in Section~\ref{sec:healthy_person}, we remove the tumor branch score, when K-CROSS evaluates the quality of neuroimage in healthy cases. We also observe that K-CROSS averagely surpasses DIST and LPIP (SOTA for natural image) by 7.8\% and 16.5\%, respectively, which proves that K-CROSS is built upon the basis of MRI principle instead of only on the natural image level. From this ablation study, we demonstrate that the performance of K-CROSS ($\mathcal{L}_{stru}$+$\mathcal{L}_{freq}$) is stable across several MRI datasets, with the potential to serve as a generic measure for evaluating the quality of the synthesized MRI.

\section{Conclusion} \label{sec:conclusion}
In this paper, we proposed a new metric K-CROSS to evaluate the performance of medical images synthesized, which is based on the principle of magnetic resonance imaging. To improve the reconstruction capability during K-CROSS training, a complex U-Net was developed. As for training a learning-based full IQA metric, we further constructed a large-scale multi-modal neuroimaging perceptual similarity (NIRPS) dataset. The experimental results indicate that K-CROSS is a useful indicator for evaluating the quality of the medical data generated. However, our method heavily relies on deep learning-based techniques but without directly injecting the knowledge of radiologists into K-CROSS. In the future, K-CROSS should combine causal inference methods to improve interpretability.

\section*{Acknowledgment}
This work is partially supported by the National Key R\&D Program of China (Grant NO. 2022YFF1202903) and the National Natural Science Foundation of China (Grant NO. 62122035,  61972188, and 62206122). 


\printcredits

\bibliographystyle{cas-model2-names}

\bibliography{cas-refs}

\bio{authors/xie1}
Guoyang Xie received the Bachelor and MPhil Degrees from University of Electronic Science and Technology of China, Hong Kong University of Science and Technology in 2009 and 2013, respectively. He is pursuing the PhD degree from University of Surrey. Prior to that, he was the Principle Perception Algorithm Engineer in Baidu and GAC, respectively. His research interests include anomaly detection, medical imaging, neural architecture search, and federated learning. \\ 
\endbio 

\vskip40pt

\bio{authors/jinbao-wang1}
Jinbao Wang received his Ph.D. degree from the University of Chinese Academy of Sciences (UCAS) in 2019. He is currently an Assistant Professor at the National Engineering Laboratory for Big Data System Computing, Shenzhen University, Shenzhen, China. His research interests include digital human modeling and driving, image anomaly detection, computer vision, and machine learning. \\
\endbio 

\vskip40pt

\bio{authors/huang}
Yawen Huang received the M.Sc. and Ph.D. degrees from the Department of Electronic and Electrical Engineering, The University of Sheffield, Sheffield, U.K., in 2015 and 2018, respectively. She is currently a Senior Scientist of Tencent Jarvis Laboratory, Shenzhen, China. Her research interests include computer vision, machine learning, medical imaging, deep learning, and practical AI for computer-aided diagnosis.
\endbio

\vskip40pt

\bio{authors/jiayi-lyu1}
Lyu Jiayi, born in 1999, graduated in 2021 from Capital Normal University with a bachelor's degree in computer science and technology. She is now pursuing a Ph.D. in computer applications at the School of Engineering Science, Chinese Academy of Sciences.\\
\endbio

\vskip40pt

\bio{authors/feng-zheng}
Feng Zheng (Member, IEEE) received the Ph.D. degree from The University of Sheffield, Sheffield, U.K., in 2017. He is currently an Assistant Professor with the Department of Computer Science and Engineering, Southern University of Science and Technology, Shenzhen, China. His research interests include machine learning, computer vision, and human-computer interaction.
\endbio

\vskip40pt

\bio{authors/yefeng-zheng}
Yefeng Zheng (Fellow, IEEE) received the B.E. and M.E. degrees from Tsinghua University, Beijing, in 1998 and 2001, respectively, and the Ph.D. degree from the University of Maryland, College Park, MD, USA, in 2005. After graduation, he joined Siemens Corporate Research, Princeton, NJ, USA. He is currently the Director and the Distinguished Scientist of Tencent Jarvis Laboratory, Shenzhen, China, leading the company’s initiative on Medical AI. His research interests include medical image analysis, graph data mining, and deep learning. Dr. Zheng is a fellow of the American Institute for Medical and Biological Engineering (AIMBE).
\endbio

\vskip40pt

\bio{authors/yaochu-jin}
Yaochu Jin (Fellow, IEEE) received the B.Sc., M.Sc., and Ph.D. degrees from Zhejiang University, Hangzhou, China, in 1988, 1991, and 1996, respectively, and the Dr.-Ing. degree from Ruhr University Bochum, Germany, in 2001. He is presently Chair Professor of AI, Head of the Trustworthy and General AI Lab, School of Engineering, Westlake University, Hangzhou, China. Prior to that, he was Alexander von Humboldt Professor for Artificial Intelligence endowed by the German Federal Ministry of Education and Research, with the Faculty of Technology, Bielefeld University, Germany from 2021 to 2023, and Surrey Distinguished Chair, Professor in Computational Intelligence, Department of Computer Science, University of Surrey, Guildford, U.K. from 2010 to 2021.  He was also “Finland Distinguished Professor” with University of Jyväskylä, Finland, and “Changjiang Distinguished Visiting Professor” with the Northeastern University, China from 2015 to 2017. His main research interests include evolutionary optimization, evolutionary machine learning, trustworthy AI, and evolutionary developmental AI. Prof Jin is the President of the IEEE Computational Intelligence Society and Editor-in-Chief of Complex \& Intelligent Systems. He was named by the Web of Science as “a Highly Cited Researcher” from 2019 to 2023 consecutively. He is a Member of Academia Europaea.

\endbio





\end{document}